# Hidden variables and hidden time in quantum theory

Kurakin P. V.
(Keldysh Institute of Applied Mathematics, Russian Academy of Sciences)

Bell's theorem proves only that hidden variables evolving in true physical time can't exist; still the theorem's meaning is usually interpreted intolerably wide. The concept of hidden time (and, in general, hidden space-time) is introduced. Such concept provides a whole new class of physical theories, fully compatible with current knowledge, but giving new tremendous possibilities. Those theories do not violate Bell's theorem.

## 1. Introduction

(a) The problem of hidden variables is historically connected to a *gedankenexperiment* proposed by Einstein, Podolsky and Rosen [1], when trying to revise some positions of quantum mechanics.

EPR - experiment (by authors' last names) in its simplest case looks as follows. Some quantum system is divided into two subsystems, while some *summary* physical quantity is preserved. Say, let it be angular momentum (spin). Then subsystems move arbitrary far from each other. If one measures the preserved quantity for some one subsystem then - at appropriate measurement basis supplied - the result of analogous measurement for paired subsystem is pre-determined.

Professor John G. Cramer of Washington University (Seattle, USA) offers, as I see it, the best description of EPR paradox and corresponding experiment in its modern view [2]:

"This kind of non-locality is demonstrated by the Freedman-Clauser experiment (1972)... Here excited calcium atoms undergo a 0+ to 1- to 0+ atomic cascade and provide a pair of photons, assumed to be emitted back-to-back, which are in a relative L=0 angular momentum state. Because of angular momentum conservation these photons are required to have identical helicities or linear combination of helicities, i.e., they must be in identical states of circular or linear polarization. For this reason the SV of the two photon system permits the photons to be in any polarization state, provided only that both are in the same state. Experimentally this means that if the photons are transmitted through perfect polarizing filters before detection, they must be transmitted with 100% probability if the polarizations of the filters select matching states and 0% if the filters select orthogonal states, no matter what orientation or polarization selectivity the filters have.

The Freedman-Clauser (FC) experiment employs linear polarizing filters and measures the coincident transmission yield of the two photon detectors when the principal axes of the two filters are set at angles $Q_A$ and $Q_B$, which are varied independently. Quantum mechanics predicts that the experimentally observed yield will depend only on the relative angle $Q_{rel} = Q_A$ - $Q_B$ between the two principal axes, and further that for ideal filters the yield will have the normalized angular dependence:

$$R[Q_{rel}] = Cos^2(Q_{rel}) \quad (1)"$$

(b) Such correlations in quantum mechanical predictions, from viewpoint of authors of [1], look very suspicious. This is why authors of [1] supposed the following: these predictions should be wrong or quantum mechanics in incomplete theory. The Freedman - Clauser experiment demonstrates that such correlations do exist indeed. So, quantum mechanics is correct.

The notion of "completeness" of a theory, as introduced by A. Einstein, is complex enough, and we shall not deal it. True physical problem, in my opinion, is that standard quantum mechanics does not *explain* in any way quantum phenomena instead it offers *predicted probability amplitudes* for elementary processes. Quantum systems do posses any trajectories. So, it is unclear *how exactly* a quantum transition from initial state to a final one is accomplished.

The theory does not explain in any way, *why* and *как* spins of arbitrary remote particles turn out to be correlated. It is very important here to understand that it's fundamentally incorrect to imagine, as if correlated values do "exist" just immediately after the parts of system depart from each other. Later we'll give a detailed explanation of this by Prof. J. Cramer. True problem here is that each subsystem gets its own value of preserved quantity at the *detection instant* only.

(c) Unlocal correlations like in FC experiment, *do not contradict* to principles of special relativity, but, as Roger Penrose states, [3], they contradict to the *spirit* of relativity.

When trying to *explain* such unlocal correlation in local way, and thus trying to *describe explicitly* any quantum system transition, A. Einstein introduced a hypothesis of «hidden variables». Some smooth and determined evolution of hidden variables should, as A. Einstein supposed, give "at output" the observed values of physical quantities.

As is commonly assumed, Bell's theorem experiments (see below) deny the possibility of hidden variables to exist in the Nature.

## 2. Bell's theorem

(a) Here's the formulation of Bell's theorem according to classical paper [4]. I suppose that I should retell the formulation in many details, since a reviewer of one respected Russian physical magazine stated that I don't catch the essence and formulation of the theorem.

Particularly, that reviewer thinks that:

"Authors' remark that information on particles and correlated quantity is, according to [4], localized in particles themselves, is not valid. Description in [4] is not in language of particles, such a notion can not even be strictly defined".

So, I cite:

"Consider an ensemble of correlated pairs of particles moving so that one enters apparatus $\mathbf{I_a}$, and the other apparatus $\mathbf{I_b}$, where **a** and **b** are adjustable apparatus parameters. In each В каждом apparatus a particle must select one of two channels labeled +1 and -1. Let the result of these selections be represented by A(**a**) and B(**b**), each of which equals ± 1 according as the first or the second channel is selected.

Suppose now that a statistical correlation of A(**a**) and B(**b**) is due to information carried by and localized within each particle, and that at some time in the past the particles constituting one pair were in contact and communication regarding *this* information".

Next, authors of [4] formulate what hidden variables theories mean in generalized view.

"The information [on correlation], which emphatically is not quantum mechanical [now that quantum information theory is intensively developed we can disagree], is [in some way] part of the content of a set of hidden variables, denoted collectively by **λ**;. The results of the two selections are then to be deterministic functions A(**a, λ**) and B(**b, λ**). Locality reasonably requires A(**a, λ**) to be independent of the parameter **b** and B(**b, λ**) to be likewise independent of **a**, since the two selections may occur at an arbitrarily great distance from each other.

Finally, since the pair of particles is generally emitted by a source in a manner physically independent of the adjustable parameters **a** and **b**, we assume that the normalized probability distribution ρ(**λ**) characterizing the ensemble is independent of **a** and **b**."

Using described assumptions, authors derive some row of inequalities, applied to statistical parameters of ensemble of measurements over correlated particles. These inequalities are referred to as Bell's inequalities; their exact view is unessential for us here. Being true in an experiment, such inequalities *do not contradict* existence of hidden variables, though it should not *prove* their existence. On the other side, being violated in an experiment, the inequalities strictly deny hidden variables.

But the inequalities are violated in that experiment.

(b) A refined experiment was proposed and executed in [5], which is classics now as well. Main innovation here, as compared to previous experiment, is to use time-varying parameters **a** and **b**, which are main players in Bell's theorem. Main axes of used optical crystals were switched randomly each 10 ns, while the light needed 40 ns to pass from one detector to another. Thus the possibility for detectors to exchange any signals was excluded. Correlation (1) was true, while Bell's inequalities were violated as well.

## 3. A fundamental procedure to measure physical time

(a) In my opinion, main omission of the theorem is implicit usage of intuitive notion of physical time as some abstract and uniform flow of "something", which contains all physical events. I pose that this implicit assumption is in contradiction to special relativity and quantum nature of elementary events.

Special relativity *explicitly* says that the distance is measured by some "ruler", i.e. some solid-body linear object. But *what* is time measured *by*? A. Einstein introduced a notion of "clock", but he didn't define it strictly.

In classical domain we can think of a clock as any, rather stable oscillatory process. And, since we *always* deal with dissipative systems, we always deal with dissipative oscillators, which use outer inflow of energy. Say, usual wall clock use gravitational energy of weight, while the watch on our arms uses spring energy of electrical battery energy.

If we measure rather large (classical) times, we can distinguish some partial periods of time. In other words, we can detect incomplete turns of clock's "hand". The question here is: which is accuracy of these measurements? In other words, how small can be small parts of full turn of clock's hand, such that we can distinguish?
Since every tiny turn of clock's "hand" is supported by supply of energy, the answer to the latter question becomes evident. The energy is quantized, thus time measurement is quantized as well. So, in each experiment elapsed time is measured by the number of absorbed quanta.

When normalized by some factor depending on definite experiment's configuration, the time elapsed in the point <u>is</u> the number of energy quanta absorbed in that point**.**

It is important for us to accept that the number of absorbed quanta is not simply *one possible* way to measure time; it is fundamentally *the single* way. All the others are deduced to this one.

(b) I propose to introduce into physical theory some fundamental procedure of detecting (measuring) of physical time, something like what A. Einstein did when he introduced a fundamental procedure to synchronize remote clocks. In other words, I propose to think of not only simultaneity, but time itself, in operational terms.

Let us assume a simple, but definite process. Let us determine what it takes for a photon to move from its source to its detector. In other words, we shall describe how *in principle* the time of photon's "flight" is detected, and, thus, which is *experimental* sense of "speed of light" notion.

The general schematic idea of time detection procedure is at Fig. 3-1. Exactly in the middle of line between source atom $\mathbf{S}_1$ and detecting atom $\mathbf{D}_1$ there is one more light source $\mathbf{S}_0$, which we call "a button switch".

```
                                                   (S2)
                                                     |
                                                     |
                                                     |
         |<----------  (S0)  ------------>|_V_
    (S1) |------------------------------------->|(D1)
```

*Fig. 3-1. Fundamental idealized procedure of detecting physical time*

Let the source-"button" simultaneously radiates 2 light signals in opposite directions. These signals come to the source and to the detector *simultaneously* (this simultaneity takes place in the referential frames where all sources and detectors are at rest). When the signal comes to the source, the latter emits a photon.

On the other side the "valve"; at right side is opened *at the same instant*. The valve shadows the detector form laser $\mathbf{S}_2$ light. The laser is adjusted to the frequency of atomic transition. Thus the detector starts getting photons from the laser.

Let us assume that all considered photons have the same energy. Let us also admit that after absorption of each photon we forcedly relax the detecting atom to its lower level (in some way).

Please note, that *in principle* we are able to distinguish a photon from the laser $\mathbf{S}_2$ from a photon from our atomic source $\mathbf{S}_1$. The reason here is that the detecting atom gets some recoil at absorption of a photon. Though confined by uncertainty relation, we are able to determine, whether momentum recoil is horizontal or vertical. The former case means the photon is from atom $\mathbf{S}_1$, while the latter means that the photon is from laser $\mathbf{S}_2$.

So, in such a way the detector starts counting photons from the laser, which begins *simultaneously* to the photon departure from the source atom. This counting stops as soon as the detecting atom gets a photon from the source atom. The "flight" time for that photon is the number of photons received from the laser up to now.

It is important to note that proposed technique is not tightly connected to detecting atom $\mathbf{D}_1$. The counter of quanta from laser can be placed in arbitrary place. But in that case the detector $\mathbf{D}_1$ should re-emit the photon from the left (from our source atom) and "redirect" it onto introduced second detector, so that the latter to know when to stop counting laser photons. This, naturally, means that we should introduce some correction into our calculations. This correction takes into account the time needed for a photon to move from $\mathbf{D}_1$ to the second detector.

(c) As far as i know Leon Brillouin was the first to propose the usage of stable laser light to measure physical time. He did so in his book *Relativity reexamined* [6].

But unfortunately L. Brillouin did not outline which is time quantum in his clock. I suppose he assumed the period of light wave to be such a quantum. But this implies *classical* character of light absorption, described by classical Maxwell equations. When essentially *quantum* character of light absorption is taken into account, such a definition for time quantum is not valid.

## 4. The concept of "hidden" time

Now we have enough initial data to clearly see and formulate main narrow bottle neck in Bell's theorem, which bans hidden variables. The matter is as follows. The assumption "distribution **ρ(λ)**, which characterizes the ensemble, does not depend on **a** and **b**, though seems to be physically evident, actually bans only some very constrained class of physical theories. As I tried to argue above, physical time is not some abstract and uniform flow of "something", where we "put in" all elementary events. Instead, physical time (space - time, to be fully correct) itself consists of these events. Physical time is measured by amount of such events. So, physical time is discrete since those elementary events are discrete.

So it is very lawful to introduce a set of variables **λ** into the theory, which are not physically detected quantities (in this sense they are true "hidden variables"), they exist in theory's *mathematical apparatus* only. Such variables evolve in so called "hidden time", which is not equivalent to physical time as well. Hidden time is *mathematical* notion only. Then, elementary events (like photon absorption by an atom) are "sewing" points between hidden time and physical time.

In a theory of *such a class* we can lawfully use any signals between sources and detectors that they exchange in *hidden time*, that is, "between" acts of radiation and absorption. Since the propagation of such signals take place not in physical time, it is senseless to talk of their "speed". If *such* hidden signals make a number of passes back and forth between sources and detectors, then, obviously, their statistical distribution can lawfully depend upon detectors' adjusted parameters **a** and **b**, which does not contradict neither locality (or causality) principle, nor confined velocity of *physical* signals.

## 5. One possible idealized model of radiative phenomena

(a) Let us first assume one photon case. The radiation and absorption happens as a result of three passes of signals from the source to the detector and backwards (Fig. 5-1):

($D_2$)<-----------------(S)------------------------->($D_1$)
($D_2$)~~~~~~~~~~~> (S) <~~~~~~~~~~~~~~~~ ($D_1$)
($D_2$)< - - - - - - - - -(S) ++++++++++++++++>($D_1$)

*Fig 5-1. Possible idealized model of radiation phenomena*

- first, the source **S** sends so called *search signals* to all possible space directions (note that it happens in hidden time);
- all potential detectors, when receiving a search signal, send backward (to the source) *query signals* to get the photon;
- The single detecting atom gets the *confirmation signal*, all others get *refuse signal.*

(b) At Fig. 5-1 all looks so as if the source **S** itself makes the lottery. But we must recall that general picture is 3-dimensional in space. Both search signals and query signals move with expansion, they "divide" in each node of space grid (Fig. 5-2). This is why *query signals* (they propagate exactly the same ribs as *search signals*, but in opposite direction) from all possible detectors compete at each space node. The *whole Universe* makes the choice! And this happens in hidden time.

Fig. 5-2 gives one possible (idealized) idea of how search signals and query signals can move. Competition of query signals occurs at each node of hidden space - time, if only this node has already let search signals pass through it before. The rules of competition are very simple: a single incoming query signal survives at each node. After this the winner signal copies itself into all "underlying" (i.e. into ribs containing incoming search signals).

```
+     +     +     +
 \   / \v  ~ \v  /
  \ /   \v~   \v/
   o     o     o
    \  v/ \v  /
     \v/   \v/
      o     o
       \   /
        \ /
         +
```

*Fig. 5-2. Competition of detectors; a single query signal survives at each node*

*"+" - atom\electron (source or potential detector)*
*"o" - space node in the vacuum*
*"\" or "/" - search signal (propagates from the source)*
*"v" - query signal (propagates from a detector)*
*"~" - some incoming query signal has lost the lottery*

Regardless of how much dynamics of signals moving back and forth is complicated, the rule "only single incoming query survives at each node" *guarantees* that finally only a single linear path from the source to winner detector will stay. That detector gets the “prize”, i.e., the emitted photon.

Let me note, that a query signal, when come to a node, can "wait" for other competition participants to come for arbitrary long period of hidden time. No query signal needs to take query to be in time. It can not be "late" anyway. The reason here is that the hidden time duration of signal's propagation is not in any way connected to physical time duration of photon's motion (see below suggested mechanism of "sewing" of physical time and hidden time).

(c) It's fun that Fig. 5-2 very much like R. Feynman's formulation of quantum mechanics, where probability amplitudes are calculated as a sum of some integrals along all possible paths for a particle to make transition from its initial state to a final one.
In the idealized model at Fig. 5-2 search signals reach each possible detector by all possible ways.

(d) Let us make some additional assumptions:

- Let us assume that "hidden time "'s simultaneity notion is Newtonian; i.e., it is absolute for the whole space.
- Let us assume that all signals propagate with one and the same speed in "hidden time".

- Let us assume that a detector organizes a *queue* of *search signals* that have come to it (Fig. 5-3).
- Each *search signal* in a queue is "developed" by the detector (i.e., the detector sends a corresponding *query signal*) only after finishing of previous signal development. This means that a detector waits for confirmation\refuse message from the source, and after that only it develops the next search signal in the queue –Fig. 5-3.

These rules are necessary to "sew" hidden time and physical time. Recall the idealized physical time detecting procedure, described above. While a search signal propagates from a source to a detector, the latter accepts search signals from the laser and puts them into the queue. "When" the source's search signal comes, it is put into the queue as well.

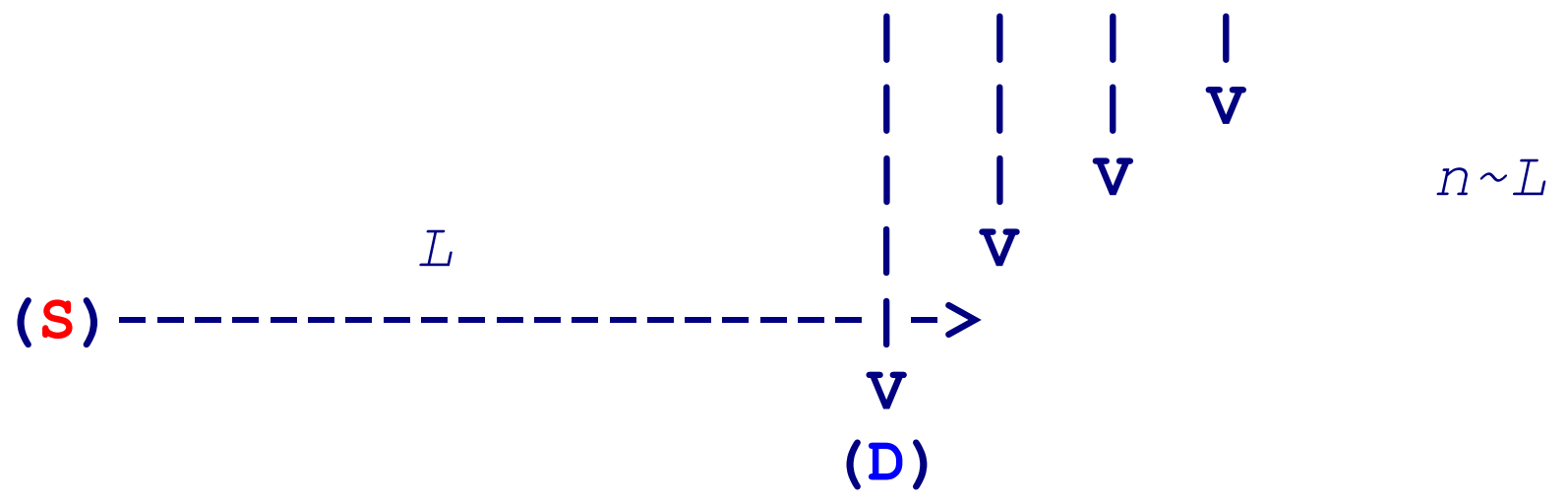

<u>*Fig. 5-3. Photon's flight time is proportional to the distance from a source to a detector.*</u>

It is clearly seen from configuration geometry at (Fig. 5-3), that number of signals in the queue *in front of* our signal (i.e. the signal from our source) is proportional to the distance from the source to the detector. It so since, at first, putting signals into the queue began *simultaneously* to departure of search signal from our source, and, at second, all query signals propagate at equal velocities (in hidden time, naturally).

Now recall rule (7). Absorption of "our" photon will happen (in physical time) will happen *after* the detector absorbs such a number of laser photons, which is obviously proportional to the distance from the source to the detector.

(e) From hidden time concept view EPR - phenomenon in its essence is described in the same manner as Fig. 5-1. "Coordination" of spins (polarizations) of two photons happens when query signals from both detectors meet at radiation point.

There is a serious unexpected trick here. Which exactly is the point in <u>space-time</u> where "coordination" of polarization occurs? In single photon case the question sounds in this way: in which space-time point the ultimate choice of final detector happens? Let us wait a little to consider this question.

## 6. "Hidden time" and TIQM by John Cramer

(a) As is known, Maxwell equations for electromagnetic field allow solution not only in the form of so called *retarded* waves: $A(r, t) \sim f(t - x/c)$, but also in the form of *advanced* waves: $A(r, t) \sim f(t + x/c)$. Advanced waves, contrary to retarded waves, are assumed to be a solution with no physical sense, because they correspond to backward in time propagation of signals (from future). So, advanced wave solutions are normally omitted. But in 1945 R. Feynman and J. Wheeler suggested that signals from future can be quite physical, provided that their interference should be correctly considered.

R. Feynman and J. Wheeler needed such a hypothesis to solve the paradox of infinite energy of self - interaction of radiating electron. They rejected this idea later.

Prof. John Cramer used Feynman-Wheeler's idea in an unusual way [2]. Not pretending to create some new theory, John Cramer built a kind of *explaining concept* of quantum radiative phenomena. According to John Cramer, this concept is only an interpretation of quantum mechanics, not new theory. This concept lets strong simplification of *understanding* weird quantum paradoxes, including EPR paradox.

John Cramer uses his transactional interpretation of quantum mechanics (TIQM) in teaching quantum mechanics in his University.

(b) Main idea of TIQM is that each elementary radiation act can be *imagined* as a kind of "simultaneous" emergence of direct (retarded) and backward (advanced) waves in space-time between the source and the detector. J. Cramer refers to this as "out-of-time 4-dimensional description". J. Cramer calls direct (retarded) wave as "proposal wave" while the backward (advanced) wave is called "confirmation wave" (Fig. 6-1).

Amplitudes of both retarded and advanced waves are equal to 1/2 of amplitude of normal (retarded) wave's amplitude, which is actually observed in an experiment. Phases of these two waves are coordinated in such a way that the radiation is absent before radiation and after detection instants. Waves interfere in such a way that their sum is zero in those regions of space-time. At the same time the amplitude of summary wave between the two points is equal to normal full-amplitude retarded wave (the observed wave).

```
 T

 ^
 |              ~   A=(1/2-1/2)
 |             ~
 |           (D)
 |         v//
 |          //
 |         //^
 |       v//    A=(1/2+1/2)
 |        //
 |       //^
 |     (S)
 |    ~
 |   ~   A=(1/2-1/2)
--------------------------------> X
```

*Fig. 6-1 Transactional interpretation of QM by John Cramer*

As one can see John Cramer uses the same basic idea as R. Feynman and J. Wheeler. What are two waves necessary for, why can't we use single usual wave? The matter is that two waves are able to accomplish coordination of boundary conditions and explain unlocal correlations like EPR.

Note that EPR - paradox is not the single one in its sense. Some more paradoxical *gedankenexperiments* (using quantum mechanical predictions) were proposed after EPR. For instance we can point to so called negative result gedankenexperiment of Renninger (when *missing* of particle in a detector makes sense) and delayed choice gedankenexperiment by Wheeler.

Wheeler's delayed choice gedankenexperiment is a modification of classical two - slit experiment. A particle (say, an electron) can fall onto photographic emulsion as usual. The trick in Wheeler's gedankenexperiment is that we can remove the emulsion from the electron's path before the electron hits this emulsion but after the instance when electron "should" pass the screen with two slits. The electron hits, in this case, one of two highly collimated detectors. They are collimated so that if an electrons hits one of them, this means the electron passed exactly one definite slit. In other words, each detector gets an electron from some definite slit only.

Wheeler's gedankenexperiment yields a paradoxical situation which contradicts to common sense. We are able to make an electron pass through some definite slit *after* it "ought" to pass through both. It's unconceivable *when* an electron makes its decision.

From TIQM viewpoint there's no any paradox. One of many possible transactions between electron source and one of possible detectors just realizes, and nothing more. Wheeler himself offers the following formulation: "no phenomenon is a phenomenon until it is an observed phenomenon". I.e., it is senseless to state that the electron "has passed already" through the screen with slits until it is detected. The screen and slits are only some constituents of the whole set of experiment's boundary conditions *as a whole*, in the sense of 4-dimensional picture. All this sounds very logical and beautiful but, unfortunately, it doesn't increase our understanding of *how exactly* the choice of some definite transaction happens. TIQM escapes answering these questions.

(c) Where and how is the choice done? This is a key question which, in my opinion, prevents TIQM to become a true *model* rather than an interpretation. If retarded and advanced waves do exist indeed at radiation instant, this means (say, in EPR phenomenon) that each of radiated photons possesses some definite polarization *immediately after radiation*. In other words, the collapse of photons' wave function occurs not at detection instants but at radiation instant. Such a hypothesis is known as Furry modification.

Prof. John Cramer explains why this suggestion is incorrect:

"We can... simulate the Furry modification within the FC experiment by placing near the source an additional pair of aligned linear polarizing filters which are rapidly and randomly changed. By this mechanism each pair of photons emerging from the source will be placed in definite and identical but sequentially random states of linear polarization as the photons are transmitted through these filters near the source...
The QM prediction for this case can easily be obtained by calculating the predicted rate of two-photon detection for a particular orientation angle (phi) of the randomizing filters and then averaging over all possible values of (phi). The result of this calculation is:

$$R_f[q_{rel}] = (1/8)[1 + 2\,Cos^2(q_{rel})] \quad (2)"$$

(d) The "hidden time" concept uses the same basic idea as TIQM: several passes of waves back and forth. But the propagation of waves is moved from physical time into hidden one. This is why the necessity of unclear "backward in time" propagation goes. Hidden time is a very simple and obvious thing: it flows forward only.
Besides, hidden time concept suggests the choice to happen in hidden time as well. And, which is essential, the choice instant corresponds to no one physical instant. This helps the idea of many passes of waves to get free of unwanted consequences described above, when TIQM is considered as "true" model.

## 7. "Hidden time" and "theory of elementary waves" by Lewis E. Little

(a) Prof. Lewis E. Little proposed one more very relative approach, based on idea of backward signals from detectors to the source. The approach is known as "theory of elementary waves" (TEW) [8]. Prof. Lewis' idea is quite unexpected - he proposes the signals from detector to propagate in normal physical time rather than in "hidden" time of "from future". According to his concept, true-time waves propagate from a possible detector to the source by all possible ways and then they interfere at the source location. Waves from one detector are coherent, but it is not so for waves from different detectors. The source gets signals from all possible detectors and generates a particle in some definite state according to calling detector with probability proportional to square of signal amplitude, which corresponds to standard formulation of QM.

(b) In short, my critic of TEW is as follows.

- TEW assumes that a particle moves in true physical time from the source to the detector. This implies that the *choice* of final state physically exists at radiation instant. But, as we showed in previous section citing Prof. John G. Cramer, such an assumption, at least in entangled photons case, is incorrect because it does not correlate to experiment. Making choice in hidden time solves this problem in correct way.
- In standard 2-slits experiment the particle, according to TEW, passes through some definite slit (no matter which exactly) in true physical time. R. Feynman gave classical consideration of why this can not be true. If a particles passes through some definite slit, we can "catch" it in the vicinity of slit. And this inevitably leads to destruction of interference picture at the screen. But, according to TEW, such a "catching" provides no reasons to destroy interference. The description becomes correct if we say that the particle moves in hidden time.
- Waves in true physical time are good indeed to describe experiments with static experiments. But the description of dynamical experiments like Wheeler's delayed choice or like [5] becomes not evident. Hidden time signals do solve this problem.
- It very rational indeed to interpret the wave function of a single particle as a true physical wave (regardless of is this wave from a source to a detector or quite opposite). But it is well known that 2-particles quantum amplitudes like $\psi(x_1, x_2) = \psi_1(x_1)\,\psi_2(x_2) \pm \psi_1(x_2)\,\psi_2(x_1)$ , can hardly be interpreted as physical wave. It is so because such a wave should be correlated not to some definite but single point in space-time, but to 2 (or even more) points. Hidden time signals are by themselves not the same as wave function, but they are principally able to accomplish distributed calculation of such amplitudes.

(c) To be true I am not ultimately sure to deny true-time waves. It's better to say that I am interested in much more intensive discussion of these ideas by scientific society. Together with TEW and TIQM, the concept of hidden time signals constitutes now the whole direction of thought, using "backward" signals from detectors to sources. So, I think, the basic idea is quite mature for quantum physicists to discuss it vast.

## 8. Conclusion

(a) Let me repeat the main idea of this paper: Bell's theorem denies some definite class of hidden variables theories, and not the general case. Theories with variables evolving in hidden time are not in the domain of Bell's theorem applicability.

(b) The concept of hidden time lets great possibilities, which are impossible in standard theory. In particular, the proposed concept lets to make a constructive approach the united nature of electrical charge and vacuum.

In this approach, a charge is what generates search signals, query signals, refuse signals and confirmation signals. Besides, electrical charge accomplishes lottery. Introduced vacuum nodes accomplish local lotteries for incoming query signals as well. The difference between charges ("true" space nodes) and vacuum nodes (nodes of hidden space) is that vacuum do not *generate* any signals, but only *conduct* them. i.e. let the signals pass through. I suppose that it is possible to construct vacuum (hidden) node in such a way, that it is composed of 2 "true" nodes with different *signs*. I.e., I suppose that vacuum node is composed of 1 virtual electron and 1 virtual positron.

I should also note that the very essence of proposed model using many paths for search signals looks very much like R. Feynman's formulation of QM. It makes the proposed model very much looking like truth.

As a result we get a hope to interconnect electron charge ***e***, speed of light ***c*** and Planck's constant ***h*** all together in some unified and *constructive* sense. Say, electron charge can possibly be responsible for signals generation and lotteries, light speed can be responsible for signals passing between vacuum nodes, and Planck's constant for spatial density of vacuum nodes.
In fact I suggest a program to calculate thin structure constant $\alpha = e^2 / hc$. As many quantum theory classicists like Max Planck or Richard Feynman noted, a *complete* quantum theory should be able to calculate that constant rather than to use its experimental value.